\documentclass[%
reprint,
superscriptaddress,
 amsmath,amssymb,
 aps,
prb,
]{revtex4-1}

\usepackage{graphicx}
\usepackage{dcolumn}
\usepackage{bm}

\begin{document}

\preprint{APS/123-QED}

\title{Magnetic sensitivity of the Microwave Cryogenic Sapphire Oscillator }
\author{Vincent Giordano}
 \affiliation{FEMTO-ST Institute, Time and Frequency dept.\\ Centre National de la Recherche Scientifique\\ 26 Rue de l'\'Epitaphe\\ 25030 Besan\c con Cedex FRANCE.}
 \email{giordano@femto-st.fr}
\author{Christophe Fluhr} 
\author{Benoit Dubois}%
\affiliation{%
FEMTO Engineering\\
15B avenue des Montboucons, 25030 Besan\c con Cedex
FRANCE
}%

\date{\today}

\begin{abstract}
The Cryogenic Sapphire Oscillator is today recognized for its unprecedented frequency stability, mainly coming from the exceptional physical properties of its resonator made in a high quality sapphire crystal. With these instruments, the fractional frequency measurement resolution, currently of the order of $1\times 10^{-16}$, is such that it is possible to detect very small phenomena like residual resonator environmental sensitivities. Thus,
we highlighted an unexpected magnetic sensitivity of the Cryogenic Sapphire Oscillator (CSO) at low magnetic field. The fractional frequency sensitivity has been preliminary evaluated to $1\times10^{-13}/$Gauss, making this phenomenon a potential cause of frequency stability limitation. In this paper we report the experimental data related to the magnetic sensitivity of the quasi-transverse magnetic Whispering Gallery (WGH) modes excited in  sapphire crystals differing from their paramagnetic contaminants concentration. The magnetic behavior of the WGH modes does not follow the expected theory combining the Curie law and the Zeeman effect affecting the Electron Spin Resonance of the paramagnetic ions present in the crystal. 
\end{abstract}

\maketitle


\section{\label{sec:1}Introduction}

The microwave Cryogenic Sapphire Oscillator (CSO) is currently the most stable source at short term presenting a fractional frequency stability better than $1\times 10^{-15}$ for integration times $\tau \leq 10,000$ s \cite{locke08,uffc-2016}. Numerous tests have already proved its effectiveness and robustness in a number of very demanding scientific applications \cite{watabe07-im,rsi-2012,AppPhysB-2014,takamizawa2014,abgrall2016}. The recent demonstration of a low consomption CSO \cite{cryogenics-2016} and its commercial availability \cite{www.uliss} paves the way  for its deployment in real field applications.

As the CSO presents state-of-the-art short term frational frequency stability, its individual characterisation was only recently made possible by the implementation of the three-cornered-hat method \cite{calosso2019}. By revealing the individual frequency stability of three independant CSOs,~we demonstrated that residual temperature variations arising from a non-perfect resonator stabilisation can be responsible for a part in the CSO short term frequency instability \cite{journal-physics-2016}. It is clear also from the observations that it remains some other residual environmental sensitivities, which need now to be identified and characterized. During the validation of a new CSO, we observed that its frequency is shifted when the sapphire resonator is submitted to a low value DC magnetic field. The fractional frequency sensitivity has been preliminary evaluated to $1\times10^{-13}/$Gauss, making this phenomenon a potential cause of frequency stability limitation.

In this paper we report the measurements we made to characterize the sapphire resonator magnetic sensitivity. Experimental results clearly demonstrate that this sensibility is related to the accidental paramagnetic contaminants present in the sapphire crystal. Combining the Curie law and the Zeeman effect of the paramagnetic impurities Electron Spin Resonance (ESR), a magnetic sensitivity is effectively expected. However the whispering gallery mode frequency dependence we observed does not obey these theoretical expectations.

\section{\label{sec:2}Preliminary experimental observations}

The CSO is based on a whispering gallery mode resonator made in a high-quality sapphire (Al$_{2}$O$_{3}$) monocrystal. Cooled near the liquid He temperature, a $10$~GHz resonator presents typically an unloaded Q-factor of  $1\times 10^{9}$. The sapphire resonator is stabilized at its operating point $T_{0}$ between $5$ and $10$~K where its  frequency is no longer sensible to the first order temperature fluctuations. Indeed at this \it{turnover temperature}\rm\ , the effect of dilatation and permittivity variations are compensated by the magnetic susceptibility induced by the paramagnetic impurities as Cr$^{3+}$, Fe$^{3+}$ or Mo$^{3+}$ present in very small concentration in the high purity sapphire crystal. $T_{0}$ is dependent on the exact impurity contains and thus is specific to each crystal. Eventually, the cryogenic resonator is used in transmission mode in a regular oscillator loop, and in reflection mode as the discriminator of a classical Pound servo, which stabilizes the electrical length of the sustaining loop. The power injected into the resonator is also stabilized.

\begin{figure}[h]
\center
\includegraphics[width=0.9\columnwidth]{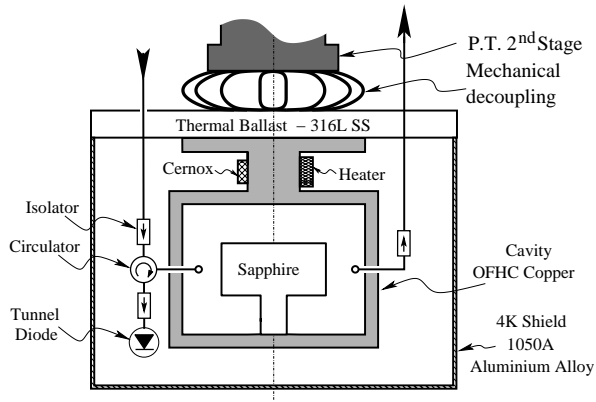}
\caption{\label{fig:fig1} The sapphire resonator mounted inside the CSO cryostat.}
\end{figure}

During the 2019 year, we undergone the building of a new CSO based on the configuration described by Fluhr \it{et al.}\rm \cite{cryogenics-2016}. The cryocooler run with a 3 kW compressor and is able to cool the sapphire resonator down to 5 K. The microwave resonator is a sapphire disk $54$~mm diameter and $30$~mm hight centered in a cylindrical copper cavity. The crystal C-axis coincides with the $z$-axis of the cylinder within $1^{\circ}$. The CSO exploites the quasi-transverse magnetic $WGH_{15,0,0}$ mode at $9.99$~GHz \cite{rsi10-elisa}. Two small magnetic loops couple the resonator to the external sustaining circuit. A ferrite circulator placed at a short distance of the input loop extracts the reflected signal useful for the Pound servo. The resonator ouput line is terminated by ferrite isolator. The figure \ref{fig:fig1} represents the immediate environment of the resonator thermally linked to the cryocooler second stage through soft copper braids.

We initially selected a sapphire crystal elaborated  with the Kyropoulos method \cite{mtt-2015}. This resonator, designed as PST-17-03, presents an unloaded Q-factor of $0.9\times 10^{9}$ measured at its turnover temperature $T_{0}=9.01$~K. This relative high value of $T_{0}$ departs from our preceeding realizations, for which $T_{0}$ was always found below $7$~K. Nevertheless, $9$~K is still compatible with the cryocooler operation and, providing the short term stability  is limited by the intrinsic noise of the control electronics, the Q-factor is sufficient to reach an Allan deviation (ADEV) of $1\times 10^{-15}$ at $\tau=1$~s \cite{journal-physics-2016}. 

\begin{figure}[h]
\includegraphics[width=\columnwidth]{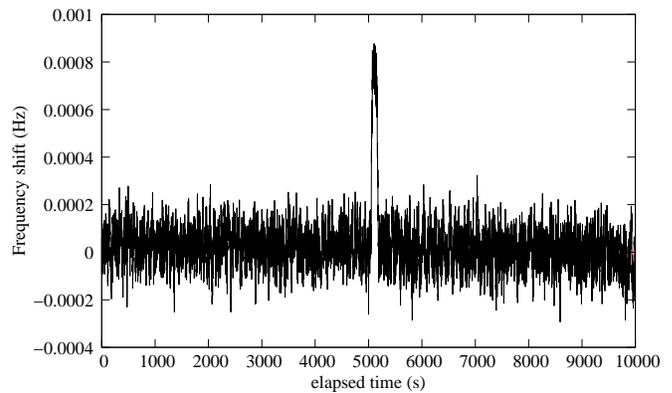}
\caption{\label{fig:fig2} Frequency shift observed when a metallic troller is passing nearby the cryostat.}
\end{figure}

Just after assembly, a preliminary measurement shows a CSO short term stability of $4 \times 10^{-15}$. Such a value is typical for a first run and can be improved by a further optimization of the resonator coupling and of the different control loops. During our first attempt to adjust the temperature and Pound servos, we observed that the CSO frequency is sensible to magnetic perturbations. The figure \ref{fig:fig2} shows the beatnote obtained by comparing the CSO under test and one of the three ultrastable CSOs of the Oscillator IMP plateform\footnote{The Oscillator IMP project funded in the frame of the french national \it{Projets d'Investivement d'Avenir}\rm\ (PIA) targets at being a facility dedicated to the measurement of noise and short-term stability of oscillators and devices in the whole radio spectrum (from MHz to THz), including microwave photonics.}. The later is implemented in an adjacent room where the temperature is stabilized within $\pm 0.5~^{\circ}$C.

At about 5,000 s, a metallic troller like those used in a mechanical workshop is moved and passes nearby the CSO's cryostat. A clear frequency shift is observed. The magnetic CSO sensitivity was later confirmed by approaching a strong permanent magnet from the cryostat. We checked that the others CSO parts, i.e. the sustaining loop and the electronic controls, are not affected by the magnet. From these preliminary observations, we roughly estimated the CSO fractional frequency sensitivity to an axial magnetic field is of the order of $1\times 10^{-13}/$Gauss. No significant frequency shift could be detected for a transversal magnetic field. \\

Inside the cryostat few components are known to be sensible to the magnetic field :

\subsection{Thermal sensor}
The observed magnetic sensitivity can not be attribuated to the Cernox\texttrademark~CX-1050~thermal sensor used to stabilize the resonator temperature at its turning point. At low magnetic field the magnetic sensitivity of this thermal sensor can be neglected \cite{brandt1999,yeager2001review}.\\

\subsection{Ferrite components}
The circulator and the isolator placed at the resonator input and output ports are commercial miniature microwave components not specialy designed for cryogenic applications. As they are based on a magnetized ferrite, they could induce a CSO magnetic sensitivity. Indeed, the resonator input and output ports are ideally terminated by $50 ~\Omega$ loads, otherwise the resonance frequency will be shifted. If $X_{G}$ and $X_{L}$ are the reactive parts of the input and output load respectively, the resonance frequency shift is given by \cite{montgomery-1947-tech}:
\begin{equation}
\label{equ:frequency-shift}
\dfrac{\Delta \nu}{\nu_{0}} = -\dfrac{\beta_{1}}{2Q_{0}}\dfrac{X_{G}}{Z_{0}}-\dfrac{\beta_{2}}{2Q_{0}} \dfrac{X_{L}}{Z_{0}}
\end{equation}

where $Z_{0}=50~\Omega$, $\beta_{1}$ and $\beta_{2}$ the input and output coupling coefficients and $Q_{0}$ the resonator unloaded quality factor.
We tested several miniature commercially available circulators and isolators. The reactances at their input port are found to be in the range $1-3~\Omega$, with a sensibility to the magnetic field of $\sim 10^{-3}~\Omega/$Gauss. With the typical resonator parameters, i.e. $\beta_{1}\approx 1$, $\beta_{2} \ll 0.1$ and $Q_{0}\approx 10^{9}$, the equation (\ref{equ:frequency-shift}) gives a magnetic field frequency sensitivity not exceeding $5\times 10^{-15}/$Gauss, which is not negligible but still to low to explain the experimental observations. 
 
\subsection{Sapphire resonator}

As it contains paramagnetic ions, the sapphire resonator is expected to be sensitive to the magnetic field. Indeed the ESR frequency of the paramagnetic dopants is dependent of the magnetic field through the Zeeman effect. A microwave resonance lying nearby the ESR frequency will be thus impacted by any change in the applied magnetic field. At a given temperature $T$, the impact of the paramagnetic dopants on the frequency $\nu$ of a $WGH$ mode is described by:
\begin{equation}
\dfrac{\nu-\nu_{0}}{\nu_{0}}= AT^{4}+\dfrac{\chi'}{2}
\label{equ:Df1}
\end{equation}

$\nu_{0}$ would be the mode frequency at $T=0$ K and in the absence of any paramagnetic dopant.
$A$ combines the temperature dependance of the dielectric constant and the thermal dilatation of the sapphire \cite{luiten96}. For the $WGH$ modes we are dealing for, $A$ is almost mode independant: $A \approx  -2.6 \times 10^{-12}$ K$^{-4}$.

$\chi'$ is the real part of the ac susceptibility for a RF magnetic field perpendicular to the crystal C-axis. It is the sum of the contributions of  all paramagnetic ion species contained into the crystal.  $\chi'$ is a multivariable function of the frequency $\nu$, the signal power, the temperature and the applied magnetic field. Let us assume a concentration $N$ of a paramagnetic ion characterized by spin $S$, a ground state zero-field-splitting $\nu_{j}$ and a spin-spin relaxation time $\tau_{2}$. At low excitation power and without any DC applied magnetic field, the real part of the susceptibility is a dispersive lonrentzian function that nulls at $\nu_{j}$ \cite{vanier-audoin-T1}:

\begin{equation}
\chi'(\nu)= \chi_{0} \dfrac{(2\pi\tau_{2})^{2}(\nu-\nu_{j}) \nu_{j}}{1+(2\pi\tau_{2})^{2}(\nu-\nu_{j})^{2}}
\label{equ:chiprim}
\end{equation}

All the temperature dependance of $\chi'$ is contained in the dc-susceptibility $\chi_{0}$, which results from the distribution of the ions on their energy levels through the effect of the thermal agitation. Assuming  $\chi_{0}$ follows the Curie law, we have:

\begin{equation}
\chi_{0} = \mu_{0} N \dfrac{g^{2} \mu_{B}^{2}  }{3 k_{B} T} S(S+1)
\label{equ:chi0C}
\end{equation}
where  $g$ is the Land\'e factor, $\mu_{B}$ the Bohr Magneton, $\mu_{0}$ is the permeability of free space and $k_{B}$ the Boltzmann constant.

\begin{figure}[h]
\includegraphics[width=\columnwidth]{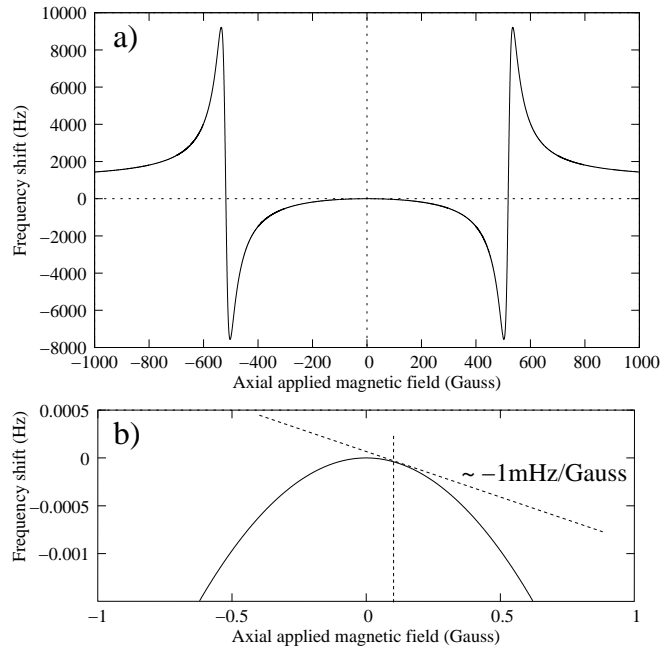}
\caption{\label{fig:fig3} Calculated frequency shift vs axial DC applied magnetic field for a $10$~GHz $WGH$ mode excited in a sapphire resonator containing $0.15$~ppm Cr$^{3+}$ and cooled at $9$~K. a) $Bz=\pm1000$~Gauss. b) $Bz=\pm 1$Gauss. The slope around $Bz=100$~mG is $\sim-1$~mHz/Gauss.}
\end{figure}

When a DC magnetic field $B_{z}$ is applied along the resonator axis, the Zeeman degeneracy of the ESR is left. The total susceptibility will be the sum of two ESR lines whose frequencies evolve linearly with $B_{z}$, i.e $\nu_{j}\pm \gamma B_{z}$ with $\gamma \sim 2.8$~MHz/Gauss.
Based on this model, the figure \ref{fig:fig3} represents the frequency shift of a $10$~GHz resonance inside a sapphire resonantor containing $0.15$~ppm of Cr$^{3+}$ ion as a function of the DC axial magnetic field.

A large frequency variation is observed when the Cr$^{3+}$ ion ESR goes through the microwave resonance, i.e. when the DC applied magnetic field is $\sim \pm 500$~Gauss. Near $B_{z}=0$, the frequency variation evolves almost quadratically. As there is no magnetic shield around the sapphire resonator, we can assume that the resonator is polarized by the Earth magnetic field of the order of few $100$~mGauss. The slope of the fractional frequency shift at this point is $\sim-1\times10^{-13}$, and thus compatible with our experimental observations. However as we will see in the next section, the experimental frequency variation substantially differs from the theoretical prediction presented in the figure \ref{fig:fig3}.\\

The sapphire resonator magnetic sensitivity has been already pointed out. It was used to tune the turnover temperature of a whispering gallery mode lying between the  Cr$^{3+}$ and Fe$^{3+}$ ions ESR frequencies \cite{kovacich97}. However, the authors insist on the fact that the apparent frequency dependence of the paramagnetic susceptibilities induced by the paramagentic impurities did not obey the expected theory. Benmessai \it{et al.}\rm\cite{prb-2009} showed how the application of an axial magnetic field on a $4$~K sapphire resonator, with a mode tuned on the Fe$^{3+}$ EPR, adds a gyrotropic component of magnetic susceptibility.
More recently, the application of a magnetic field generated by a supraconducting coil has been used to study the interaction of a two-level quantum system, i.e. the paramagnetic dopants, with the electromagnetic field of a whispering gallery mode resonator cooled down $20$~mK \cite{benmessai2013,farr2013,farr2014,goryachev2014}. In these last experiments, the environmental conditions differ greatly from our own set-up where the applied magnetic field stays below $40$ Gauss and the temperature is higher than $4$~K. 

\section{\label{sec:4}WGH Modes Magnetic sensitivity}

In order to study its magnetic sensitivity, the sapphire resonator was dismounted from the CSO and placed in a cryostat devoted to components testing. An axial magnetic field up to 36 Gauss can be applied by the means of  external Helmholtz coils. The resonator was mounted alone without any ferrite isolator and all its surounding is made with non magnetic materials.\\

The frequency of  the quasi-transverse magnetic whispering gallery modes $WGH_{m,0,0}$ has been measured with a Vector Network Analyser (VNA) referenced to a Hydrogen Maser to ensure long term stability. The $WGH_{m,0,0}$ modes separation is $\sim 570$~MHz and the $WGH_{15,0,0}$ mode frequency is $\sim 9.99$~GHz. At zero magnetic field, the ESR frequency of the Cr$^{3+}$ and Fe$^{3+}$ ions is $11.44$~GHz and $12.04$~GHz respectively.  \\

\begin{figure}[h]
\includegraphics[width=\columnwidth]{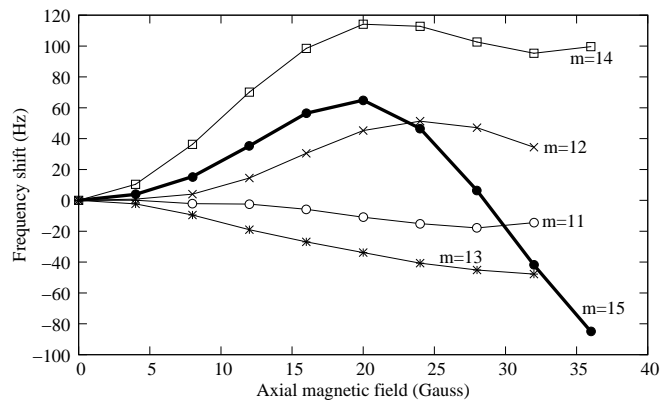}
\caption{\label{fig:fig4} Experimental frequency shift as a function of the DC applied axial magnetic field for $11 \leq m \leq 15$. In bold : the $WGH_{15,0,0}$ mode at $9.99$~GHz exploited in the CSO. }
\end{figure}

\begin{figure}[h]
\includegraphics[width=\columnwidth]{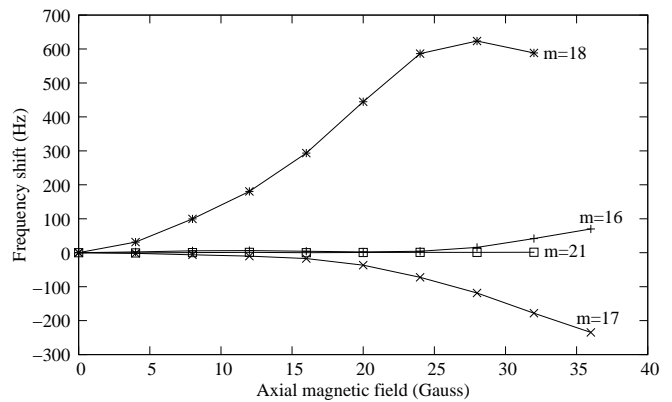}
\caption{\label{fig:fig5} Experimental frequency shift as a function of the DC applied axial magnetic field for $16 \leq m \leq 21$..}
\end{figure}

The figures \ref{fig:fig4} and \ref{fig:fig5} show the frequency shift as a function of the DC applied axial magnetic field for the modes lying nearby the ESR frequency of Cr$^{3+}$ and Fe$^{3+}$. For all the modes we followed, the frequency shift appears as an even function of the DC applied magnetic field. Thus we only present here the frequency variations for $B_{z}\geq 0$. The measurement frequency resolution depends on the mode loaded Q-factor and on the V.N.A IF bandwidth we selected. This resolution was $\pm 4$~Hz for the $WGH_{11,0,0}$ mode and better than $1$~Hz for all the following modes. \\


For an azimutal number $m<11$, no noticeable frequency shift has been observed. Above $m=18$ the $WGH$ are poorly coupled and thus difficult to observe. Nevertheless we have been able to follow the $WGH_{21,0,0}$ mode at $13.22$~GHz, which presents a very low frequency variation. It appears that none of the modes follow the expected variation described in the figure \ref{fig:fig3}. Even more surprising, the modes evolve differently with a frequency shift that can be positive or negative, while the theory predicts a negative frequency shift that should decreases when the mode frequency goes away from the ESR frequency.
The $WGH_{18,0,0}$ mode lying between the two ESR frequencies presents the larger frequency shift. This mode appears also very sensitive to the injected power due to the very fast saturation of the electron spin resonances \cite{jap-2014}.\\

From the equation \ref{equ:Df1} and making explicit only the dependence in the magnetic field, the relative frequency shift that is observed when $B_{z}$ is applied, can be written as:

\begin{equation}
\dfrac{\nu(B_{z})-\nu(0)}{\nu_{0}}=\dfrac{1}{2} \left [ \chi'(Bz)-\chi'(0) \right ]= \dfrac{\Delta \chi'(B_{z)}}{2}
\label{equ:equ5}
\end{equation}

\begin{figure}[h]
\includegraphics[width=\columnwidth]{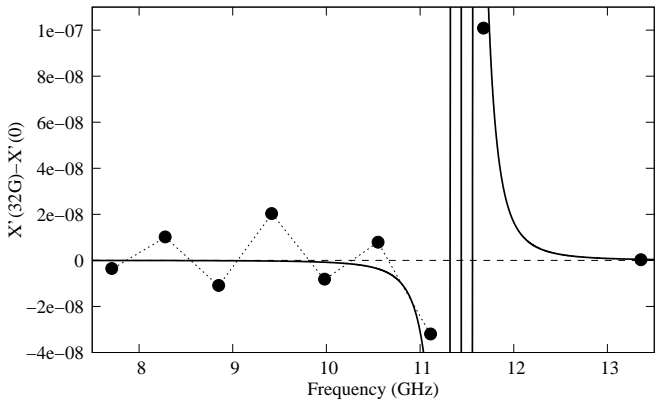}
\caption{\label{fig:fig6} Magnetic susceptibility variation when an axial magnetic field $Bz=32$~Gauss is applied. (\large$\bullet$\small) : deduced from the experimental data. Solid line: theoretical expectation.}
\end{figure}

From the collected data presented in the figure \ref{fig:fig5} and with the help of equation \ref{equ:equ5}, we calculated for each mode frequency the magnetic susceptibility variation when a $32$~ Gauss axial magnetic field is applied. The result is compared with the theoretical expectation in the figure \ref{fig:fig6}.

For the modes lying below the Cr$^{3+}$ ESR, the experimental magnetic susceptibility variation appears like an oscillating function of the frequency $\nu$ while the theory predicts a monotonous and negative function. Moreover, the experimental variation is, in absolute value, much higher than the predicted one. For the $WGH_{15,0,0}$ mode at $10$~GHz, the magnetic susceptibility added by $150$~ppb of chromium ions as calculated from the equation \ref{equ:chiprim} is $\chi'(0)=-2.4\times 10^{-7}$. Its variation when a $32$~Gauss axial magnetic field is applied is $\Delta \chi'(32 \mathrm{~Gauss})= -8.5\times10^{-9}$ while the theory predicts $-8.8\times10^{-10}$.

\section{\label{sec:5}Impact of the paramagnetic impurity concentration}
\begin{figure}[h]
\includegraphics[width=0.9\columnwidth]{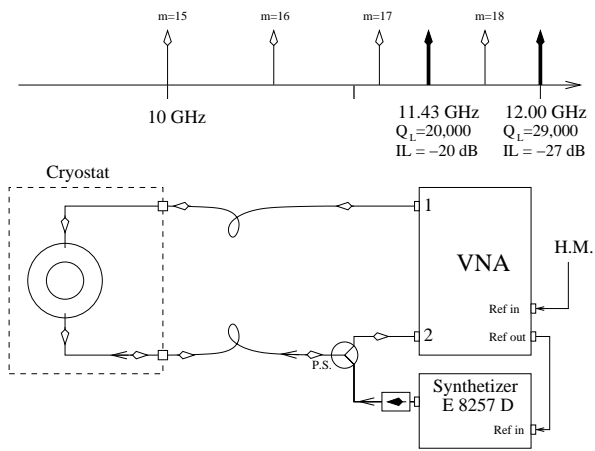}
\caption{\label{fig:fig7}Set-up used to evaluate the paramagnetic impurities concentration. Two low Q modes at $11.43$~GHz and $12.00$~GHz are excited to saturate the Cr$^{3+}$ and Fe$^{3+}$ ESR respectively.}
\end{figure}

From the preceeding experimental observations, we can reasonably conclude that the resonator magnetic sensitivity is due to the paramagnetic impurities present in the sapphire crystal. To evaluate their concentration, we adapted the method described by Mann \it{et al.}\rm \cite{Mann00}, which consists to measure the whispering gallery mode frequency shift when a strong auxiliary signal is applied to saturate the ESR transition. The set-up we used is presented in the figure \ref{fig:fig7}.\\

Around 10 GHz and apart the whispering gallery modes, the sapphire resonator presents a high density of modes with lower Q-factor. We selected two microwave resonances near the ESR frequency of Cr$^{3+}$ and Fe$^{3+}$ whose characteristics are given in the figure \ref{fig:fig7}.
An additional frequency synthetizer is used to inject into the resonator a pump signal that will saturate the ESR of the paramagnetic impurity. If the pump signal power is sufficient, the energy levels populations of the paramagnetic ion become equal and the induced susceptibility tends to zero : the pump wipe out the effect of the paramagnetic dopants. \\

Keeping all the other environmental parameters constant, the temperature sensitivity of a given whispering gallery mode can be rewritten as:

\begin{equation}
\dfrac{\nu(T)-\nu_{0}}{\nu_{0}}= AT^{4}+\dfrac{C}{T}
\end{equation}
 $C$ is the Curie coefficient, which combines all the contribution of the different paramagnetic species contained in the crystal. If the mode frequency is such that $C<0$, the mode is thermally compensated at the turnover temperature $T_{0}= \left (C/4A \right )^{1/5}$ . If one ESR is saturated by an additional signal, the mode thermal behavior will be modified. To determine the turnover temperature, the resonator temperature set-point is increased step by step. At each step and after waiting for the system stabilization the modes resonance frequencies are measured with the V.N.A. For each mode, this operation is realized the pump signal on and off.

\begin{figure}[h]
\includegraphics[width=0.85\columnwidth]{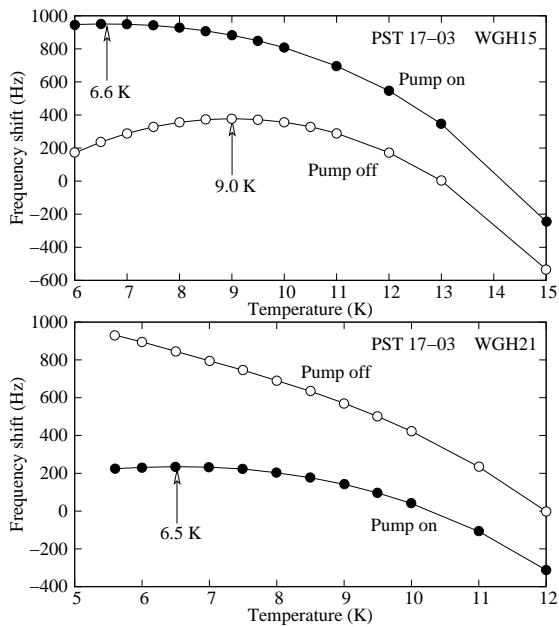}
\caption{\label{fig:fig8}Thermal behavior of the $WGH_{15,0,0}$ and $WGH_{21,0,0}$ modes with and without the pump signal at $11.43$~GHz.}
\end{figure}

 With the pump signal at $12$~GHz, i.e. very near the Fe$^{3+}$ ESR, no variation has been detected in the thermal behavior of the $WGH$ modes. Thus the concentration of Fe$^{3+}$ is very low and can be neglected. Two examples are given in the figure~\ref{fig:fig8} for a pump signal at $11.43$~GHz, which saturates the  Cr$^{3+}$ ESR.

The $WGH_{15,0,0}$ at $9.99$~GHz presents initially a turnover temperature of $9.0$~K. When the pump signal at $11.43$~GHz is applied, the susceptibility induced by Cr$^{3+}$ ions tends to zero. This mode still presents a turnover but at a lower temperature, i.e. $6.6$~K. \\
As its frequency is higher than the Cr$^{3+}$ ESR frequency,  the $WGH_{21,0,0}$ initially shows a monotonic thermal behavior without turnover. When the pump is applied a turnover appears at $6.5$~K.\\

\begin{figure}[h]
\includegraphics[width=\columnwidth]{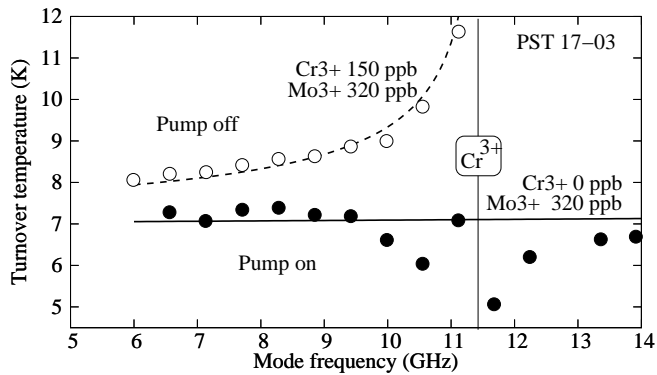}
\caption{\label{fig:fig9}Experimental turnover temperature of the $WGH_{m,0,0}$ modes: with (\large$\bullet$\small) and without (\large$\circ$\small) the pump signal at $11.43$~GHz.}
\end{figure}

The figure \ref{fig:fig9} summarizes the experimental results for the $WGH$ modes with $8\leq m\leq 22$.

With the pump signal off and starting from the lowest frequencies, the turnover temperature increases as the mode approaches the Cr$^{3+}$ ESR. The modes lying above the Cr$^{3+}$ ESR do not present a turnover point, indicating  that this ion has a predominant effect on the thermal behavior of the $WGH$ modes.
When the pump signal is on, almost all the modes presents a turnover at about $7$~K. Such a constant in the turnover temperature can be observed in the purest sapphire crystals where the Mo$^{3+}$ ion, whose ESR frequency is 165 GHz, is the predominant paramagnetic impurity \cite{luiten96}. The spread in turnover temperatures observed for the  modes near $11$~GHz could result from Cr$^{3+}$ or/and Fe$^{3+}$ residuals. We first search for the Mo$^{3+}$ ion concentration leading to a turnover temperature of $~7$~K. The solid line in the figure \ref{fig:fig9} has been obtained with  $320$ ppb of Mo$^{3+}$. In a second step we added to the model $150$~ppb of Cr$^{3+}$ ions, and obtained the dashed line shown in the figure  \ref{fig:fig9}, which represents well the experimental observations. \\

This procedure has been also applied to a second sapphire crystal coming from another manufacturer and elaborated using the Top Seeded Melt Growth (TSMG) process \cite{kawaminami2014}. This second crystal is designed as SHI-18-01 and presents for the $WGH_{15,0,0}$ mode a turnover temperature of $7$~K. The table \ref{tab:tab1} reports the paramagnetic ion concentrations determined with the above described method.

\begin{table}[h!!!!!!!!!]
\centering
\caption{\label{tab:tab1}Paramagnetic ion concentrations in the two sapphire resonators}
\begin{tabular}{lccccc}
\hline
Sapphire 		& $WGH_{15,0,0}$		&~~~~&&&	 \\
sample		& turnover temp. $T_{0}$	&& Cr$^{3+}$	& Fe$^{3+}$	&Mo$^{3+}$\\
			& (K)					&&(ppb)		&(ppb)		&(ppb)\\
\hline
PST 17-03	&	9.0				&& 150		&0	&320 \\
SHI 18-01		&	7.0				&& 30		&2	&150 \\
\hline
\end{tabular}
\end{table} 

\begin{figure}[h]
\includegraphics[width=\columnwidth]{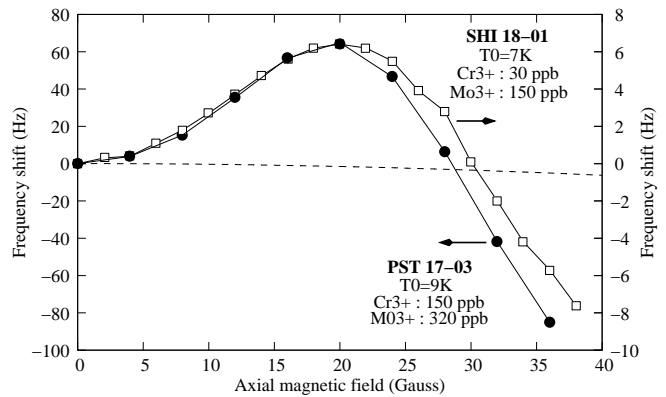}
\caption{\label{fig:fig10}$WGH_{15,0,0}$ magnetic sensitivity for the two sapphire resonators. \large{$\bullet$}\small: PST-17-03, \footnotesize{$\square$}\small: SHI 18-01. Mark well the scale difference in the frequency shifts. The dashed line is the calculated frequency shift assuming $150$~ppb of Cr$^{3+}$: the corresponding values should be read on the left scale.}
\end{figure}

The figure \ref{fig:fig10} compares the $WGH_{15,0,0}$ magnetic sensitivity for these two sapphire resonators. 

The shape of the frequency variation is the same but the SHI 18-01 resonator is almost ten times less sensible than PST 17-03. This last measurement confirms that the magnetic sensitivity is related to the quantity of paramagnetic dopants, but does not follow the classical theory. In addition, the fact that these two crystals were developed with two different growth processes and by two different manufacturers leads to the exclusion of a cause related to any inhomogeneity in the concentration of paramagnetic ions.

\section{\label{sec:5} Conclusion }
We highlighted an unexpected magnetic sensitivity of the whispering gallery modes excited in a microwave cryogenic sapphire resonator. From our experimental observations, it is clear that this magnetic sensitivity is related to the paramagnetic impurities contained inside the resonator, and especially the Cr$^{3+}$ ion. However, if the classical theory based on the Curie law describes correctly the resonator thermal behavior, it falls to explain the frequency variations observed when a DC magnetic field  is applied.  This discrepancy between the theory for isolated ions and the observed magnetic behavior is not fully understood at this time. A similar discrepancy was noted in the work of Kovacich \it{et al.}\rm\cite{kovacich97} but in different experimental conditions. To search to explain the deviations from the predicted individual ion behavior, one can invoke: i) the weak antiferromagnetic exchange coupling between distant pairs of Cr$^{3+}$ ions \cite{statz1961}, ii) the presence of other paramagnetic impurities presenting an hyperfine structure as Vanadium\cite{lambe1960,farr2013}, or $^{53}$Cr isotope \cite{terhune1961}.
These effects are second order phenomena and are generally observed in highly doped sapphire crystals in classical ESR experiments. Although our crystals are weakly contaminated, the large Q-factor of the whispering gallery mode sapphire resonator provides a sufficient resolution to reveal such a type of very weak phenomena.

\begin{acknowledgments}
The work has been realized in the frame of the ANR project Equipex Oscillator IMP. The authors would like to thank the Council of the R\'egion de Franche-Comt\'e for its support to the \it{Projets d'Investissements d'Avenir}\rm.
\end{acknowledgments}


%

\end{document}